# Spectral Properties of Quasiparticle Excitations Induced by Magnetic Moments in Superconductors

M.I. Salkola[1], A.V. Balatsky[2], and J.R. Schrieffer[3]

[1] *Department of Physics, Stanford University, Stanford, California 94305*
[2] *Theoretical Division, Los Alamos National Laboratory, Los Alamos, New Mexico 87545*
[3] *NHMFL and Department of Physics, Florida State University, Tallahassee, Florida 32310*

(October 16, 1996)

The consequences of localized, classical magnetic moments in superconductors are explored and their effect on the spectral properties of the intragap bound states is studied. Above a critical moment, a localized quasiparticle excitation in an $s$-wave superconductor is spontaneously created near a magnetic impurity, inducing a zero-temperature quantum transition. In this transition, the spin quantum number of the ground state changes from zero to $\frac{1}{2}$, while the total charge remains the same. In contrast, the spin-unpolarized ground state of a $d$-wave superconductor is found to be stable for any value of the magnetic moment when the normal-state energy spectrum possesses particle-hole symmetry. The effect of impurity scattering on the quasiparticle states is interpreted in the spirit of relevant symmetries of the clean superconductor. The results obtained by the non-self-consistent ($T$ matrix) and the self-consistent mean-field approximations are compared and qualitative agreement between the two schemes is found in the regime where the coherence length is longer than the Fermi length.

## I. INTRODUCTION

One of the most interesting phenomena in superconductors is their response to defects. They can be applied as experimental probes in studying the properties of the superconducting state. The simplest example of pair-breaking defects is magnetic impurities in singlet superconductors, where magnetic scattering disrupts pairing in the singlet channel. Pair-breaking defects are known to lead to the formation of bound quasiparticle states in conventional nodeless superconductors [1–3] and virtual bound states in gapless superconductors [4–7]. These states are localized in the neighborhood of the defects and can have a very anisotropic structure, depending on the energy-gap structure [4–7]. Quasiparticle states, localized near impurities, carry spin and are expected to modify the ground-state properties of the superconductors as well. Because experimental techniques are becoming increasingly capable of analyzing the effect of impurities on the superconductivity, it is also interesting to explore theoretically the implications of impurity scattering together with the symmetries present in the problem. Indeed, the progression from tunnel-junction spectroscopy [8] to scanning-tunneling microscopy [9–11] provides a strong evidence of inhomogeneous quasiparticle states induced by defects in superconductors.

In this paper, we revisit the problem of a localized magnetic moment interacting with a superconductor. A remarkable aspect of this interaction is the first-order zero-temperature transition which takes place in an $s$-wave superconductor as a function of the "magnetic moment", $w = JS/2$, where $S$ is the local impurity spin and $J$ is the exchange coupling. In this transition, the spin quantum number $s$ of the electronic ground state of the superconductor $|\Phi_0\rangle_w$ changes from zero for a subcritical moment $w < w_c$ to $\frac{1}{2}$ for $w > w_c$. The total spin becomes $S \pm \frac{1}{2}$ depending on the sign of the exchange interaction $J$. The first to point out the phase transition was Sakurai [12] who showed that the transition corresponds to a level crossing between two ground states as a function of the exchange coupling $J$. The level crossing occurs in a singlet superconductor between states where the impurity spin is either unscreened or partially screened. We note here that the level crossing between two ground-state energies does not occur at the point where the excitation level becomes a zero-energy excitation, since the free energy of the superconductor is determined by taking into account continuum states as well as any localized intragap states in the self-consistent solution.

We address the above problem at zero temperature by using the mean-field approximation both within the $T$ matrix formulation and the self-consistent approach which takes into account a local gap-function relaxation. We also consider a local Coulomb interaction $U$ which breaks particle-hole symmetry leading to an asymmetric spectral density for the impurity-induced quasiparticle states. Figure 1 illustrates the local effect of a magnetic moment on the low-energy spectral density in an $s$-wave superconductor. Since we limit our considerations to a classical spin, $S \gg 1$, the impurity moment cannot be screened completely by the quasiparticles. Our main results are as follows. ($i$) We show that the gross features of the impurity-induced quasiparticle states and their spec-tral properties in $s$- and $d$-wave superconductors can be qualitatively understood within the nonself-consistent $T$ matrix formalism. (ii) For an $s$-wave superconductor, we find that, above the critical moment $w_c$, the order parameter changes sign at the impurity site and that it vanishes as $w \to \infty$ [13]. In the self-consistent mean-field approximation, we find that order-parameter relaxation shifts $w_c$ downwards and that the energy of the impurity-induced bound state does not reach zero before a first-order transition between the two ground states occurs. (iii) In contrast, a $d$-wave superconductor has no quantum transition for any value of the magnetic moment when its quasiparticle spectrum in the normal state has particle-hole symmetry. The absence of the transition follows from the behavior of the impurity-induced quasiparticle states which are pinned at the chemical potential for an arbitrarily large magnetic moment. However, if particle-hole symmetry is broken or if the pairing state acquires a small $s$-wave component, the transition is again possible for a large enough moment. The $d$-wave order parameter does not appear to change sign at the impurity site. The impurity moment induces two virtual-bound states which have four-fold symmetry and extend along the nodal directions of the energy gap. (iv) Finally, a mapping to an effective theory is derived which allows the formation of the impurity band at finite impurity concentrations to be explored.

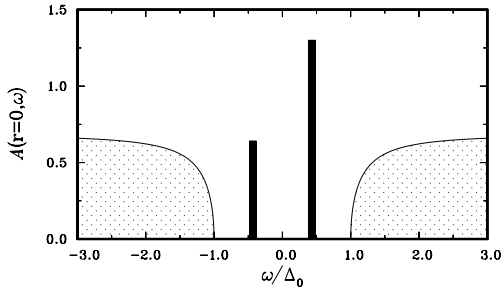

FIG. 1. The spin-unpolarized spectral density $A(\mathbf{r}=0,\omega)$ in an $s$-wave superconductor due to a magnetic moment located at $\mathbf{r}=0$ as predicted by the $T$ matrix approach. The continuum contribution is given in units of $N_F$ and the height of the intragap bound-state peaks, located at $\Omega_0 = \pm 0.43\Delta_0$, denotes their integrated spectral weights in units of $N_F\Delta_0$. Here, $N_F$ is the density of states at the Fermi energy in the normal state and $\Delta_0$ is superconducting energy gap. The magnitude of the magnetic moment is $\pi N_F w = 0.7$ and the local Coulomb interaction is $\pi N_F U = 0.4$.

The Kondo effect in an $s$-wave superconductor interacting with a quantum impurity spin was considered in a recent work [15] which found a similar type of a quantum transition from the spin-doublet state (unscreened impurity spin) to the spin-singlet state (screened impurity spin). For the antiferromagnetic coupling ($J > 0$), the two states cross at $T_K/\Delta_0 = 0.3$, whereas, for the ferromagnetic coupling ($J < 0$), the coupling constant flows to the weak coupling limit and the level crossing is absent. The situation resembles the antiferromagnetic Kondo problem in that the moment is screened. However, the crossover is replaced here by a sharp transition [15].

The physical picture of the quantum transition follows from the behavior of the impurity-induced bound state. This transition is a consequence of the instability of the spin-unpolarized ground state, because, for a large enough $w$, the energy of the impurity-induced quasiparticle excitation would fall below the chemical potential. For a weak coupling $w < w_c$, the ground state of the superconductor is a paired state of time-reversed single-particle states in the presence of the impurity scattering [16], $|\Phi_0\rangle_{w<w_c} \sim |\Phi_0\rangle$, where

$$|\Phi_0\rangle = \prod_{i>0}(u_i + v_i \psi_{i\uparrow}^\dagger \psi_{-i\downarrow}^\dagger)|0\rangle, \qquad (1)$$

and the bound state at the energy $\Omega_0$ is an intragap excited state, $|\Phi_{-1\downarrow}\rangle_{w<w_c} \sim |\Phi_{-1\downarrow}\rangle \equiv \gamma_{-1\downarrow}^\dagger |\Phi_0\rangle$, where

$$|\Phi_{-1\downarrow}\rangle = \psi_{-1\downarrow}^\dagger \prod_{i>1}(u_i + v_i \psi_{i\uparrow}^\dagger \psi_{-i\downarrow}^\dagger)|0\rangle. \qquad (2)$$

The quasiparticle operators [17] are $\gamma_{i\uparrow} = u_i\psi_{i\uparrow} - v_i\psi_{-i\downarrow}^\dagger$ and $\gamma_{-i\downarrow}^\dagger = u_i\psi_{-i\downarrow}^\dagger + v_i\psi_{i\uparrow}$; $-i$ indicates the time-reversal conjugate of $i$ and $u_i^2 + v_i^2 = 1$. As we find below, the time-reversal conjugate of this state, $|\Phi_{1\uparrow}\rangle_{w<w_c} \sim \gamma_{1\uparrow}^\dagger|\Phi_0\rangle$, does not appear in the superconducting energy gap. The processes of adding an electron with spin down to state $i = -1$ and removing an electron with spin up from state $i = 1$ in the ground state $|\Phi_0\rangle_{w<w_c}$ lead to the *same* excited state: $\psi_{-1\downarrow}^\dagger |\Phi_0\rangle = u_1|\Phi_{-1\downarrow}\rangle$ and $\psi_{1\uparrow}|\Phi_0\rangle = v_1|\Phi_{-1\downarrow}\rangle$. Thus, in a tunneling experiment, the bound state is responsible through these processes for a finite differential conductance proportional to $u_i^2$ and $v_i^2$ at energies $\omega = \pm\Omega_0$. Figure 2 elucidates schematically the mixing of particle and hole degrees of freedom in a quasiparticle state using a potential scatterer as an example. Because the particle and hole components are part of the same state, they appear symmetrically relative to the chemical potential. As $w$ increases, $\Omega_0$ approaches the chemical potential and, at the critical coupling $w_c$, it becomes a zero-energy state. At this point, the ground state of the superconductor becomes unstable. The new ground state of the superconductor becomes a paired condensate for all but the impurity bound state *with a finite single-particle amplitude at the impurity-bound state*. Thus, it has the form $|\Phi_0\rangle_{w>w_c} \sim |\Phi_{-1\downarrow}\rangle$. This unpaired ground state is a result of the competition between the pairing-condensation energy and the magnetic interaction. For a strong magnetic interaction, the gain in the magnetic energy, due to the induced spin, dominates the condensation energy. The (variational) forms (1) and

(2) allow a varying number of doubly excited quasiparticles of the pure superconductor created by the operators $\gamma_{i\uparrow}^\dagger \gamma_{-i\downarrow}^\dagger$ from $|\Phi_0\rangle$. These lead to a local suppression of the gap function $\Delta(\mathbf{r})$ and, in particular, for $w > w_c$, to a negative value of $\Delta(\mathbf{r})$ at the impurity site (see below and Ref. [14]). The spin density is spread over a volume of linear dimension of the order of the coherence length.

The plan of the paper is as follows. In Section II, the general formalism is introduced. Sections III and IV explore the consequences of a magnetic moment in $s$- and $d$-wave superconductors. Both uniform and non-uniform order parameters are considered within the $T$ matrix and self-consistent mean-field approximations. In Section V, the generalization to many impurities is outlined. Section VI contains a discussion of the number of electronic degrees of freedom associated with the impurity-induced quasiparticle states. In Appendix A, the interplay between impurity scattering and pairing symmetry is further elaborated from the quasiparticle perspective. In Appendix B, the $T$ matrix is generalized to finite impurity concentrations.

## II. FORMALISM

Our starting point in describing a localized magnetic moment in a superconductor is the lattice formulation of electrons hopping between nearest-neighbor sites and interacting via an effective two-particle interaction. The pairing interaction is assumed to be weak or at most moderate so that the mean-field approximation gives a qualitatively reliable description of the superconducting ground state and the low-energy excitations. Specifically, consider a two-dimensional lattice model where a localized magnetic moment is created by a classical spin $\mathbf{S}$ at $\mathbf{r} = 0$. The model is defined by the effective Hamiltonian $H = H_0 + H_{\text{imp}}$, where $H_0$ describes a BCS superconductor [18] and $H_{\text{imp}}$ is the contribution due to the magnetic moment. In the mean-field approximation,

$$H_0 = -\tfrac{1}{4}W \sum_{\langle \mathbf{Rr} \rangle} \Psi^\dagger(\mathbf{R}+\mathbf{r})\hat{\tau}_3 \Psi(\mathbf{R}) - \mu \sum_{\mathbf{R}} \Psi^\dagger(\mathbf{R})\hat{\tau}_3 \Psi(\mathbf{R}) - \sum_{\mathbf{Rr}} \Delta(\mathbf{R},\mathbf{r}) \Psi^\dagger(\mathbf{R}+\mathbf{r})\hat{\tau}_1 \Psi(\mathbf{R}), \qquad (3)$$

where $\langle \mathbf{Rr} \rangle$ denotes nearest-neighbor sites separated by $\mathbf{r}$, $W$ is the half bandwidth (on a square lattice), $\mu$ is the chemical potential, and $\Delta(\mathbf{R},\mathbf{r})$ is the superconducting gap function. The operator $\Psi(\mathbf{r}) = [\psi_\uparrow(\mathbf{r})\ \psi_\downarrow^\dagger(\mathbf{r})]^T$ is a two-component Gor'kov-Nambu spinor, $\hat{\tau}_\alpha$ $(\alpha = 1,2,3)$ are the Pauli matrices for particle-hole degrees of freedom, and $\hat{\tau}_0$ is the unit matrix. Given that the pairing of electrons occurs in the spin-singlet state, the superconducting order parameter (amplitude) can be expressed in the form

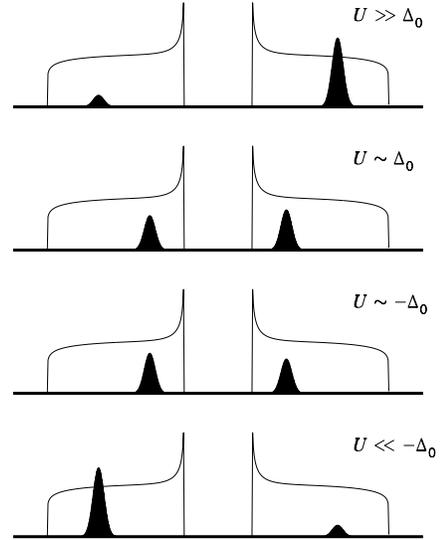

FIG. 2. A schematic evolution of the spectral density in a superconductor as a function of a local Coulomb interaction $U$. The contribution due to the impurity-induced resonance state is depicted by the shadowed peaks. For $|U| \sim \Delta_0$, the resonance has nearly equal weights in the electron and the hole components, whereas, for $|U| \gg \Delta_0$, it is mostly either electron- or holelike.

$$F(\mathbf{R},\mathbf{r}) = \tfrac{1}{2} \sum_{\sigma\nu} (i\hat{\tau}_2)_{\sigma\nu} \langle \psi_\nu(\mathbf{R}+\mathbf{r}) \psi_\sigma(\mathbf{R}) \rangle. \qquad (4)$$

The relation between the order parameter and the gap function is given by the equation

$$\Delta(\mathbf{R},\mathbf{r}) = -v(\mathbf{r}) F(\mathbf{R},\mathbf{r}). \qquad (5)$$

Here, $v(\mathbf{r})$ is the strength of the two-particle interaction, which is assumed to be instantaneous in time. Thus, the energy cutoff in the gap equation is set by the bandwidth. In a translationally invariant system, the BCS Hamiltonian reduces to

$$H_0 = \sum_{\mathbf{k}} \Psi_{\mathbf{k}}^\dagger (\epsilon_{\mathbf{k}} \hat{\tau}_3 - \Delta_{\mathbf{k}} \hat{\tau}_1) \Psi_{\mathbf{k}}. \qquad (6)$$

where $\Delta_{\mathbf{k}} = \sum_{\mathbf{r}} \Delta(\mathbf{R},\mathbf{r}) e^{-i\mathbf{k}\cdot\mathbf{r}}$ and $\Psi_{\mathbf{k}} = (\psi_{\mathbf{k}\uparrow}\ \psi_{-\mathbf{k}\downarrow}^\dagger)^T$. The fermion operators in real and momentum space are related by the unitary transformation, $\psi_\sigma(\mathbf{r}) = N^{-1/2} \sum_{\mathbf{k}} \psi_{\mathbf{k}\sigma} e^{i\mathbf{k}\cdot\mathbf{r}}$, where $N$ is the number of sites in the system. For a square lattice with the nearest-neighbor hopping, the single-particle energy relative to the chemical potential in the normal state is $\epsilon_{\mathbf{k}} = -\tfrac{1}{2} W (\cos k_x a + \cos k_y a) - \mu$; $a$ is the lattice spacing. At half filling, $\mu = 0$, the effective Hamiltonian with nonzero on-site order-parameter is symmetric under the particle-hole transformation generated by the operator $\hat{\tau}_1$: $\Psi_{\mathbf{k}} \to \Psi_{\mathbf{k}}' = \hat{\tau}_1 \Psi_{\mathbf{Q}-\mathbf{k}}$, where $\mathbf{Q} = (\pi/a, \pi/a)$.

The nature of the superconducting condensate can be established by studying the order parameter $F$ which incorporates the pairing correlations and gives an estimate for the size of the Cooper pairs [18]. (i) An on-site attraction favors $s$-wave pairing characterized by $F(\mathbf{R}, \mathbf{r} = 0) \neq 0$ and $\Delta_\mathbf{k} = \Delta_0$. As an example, consider a uniform superconductor close to half filling ($\mu \sim 0$) so that the Fermi surface is square. Let $\mathbf{n}$ be the normal to the two sheets of the Fermi surface which are most nearly normal to $\mathbf{r}$ and $r_\perp \equiv \mathbf{n} \cdot \mathbf{r} > 0$. Thus, $r_\perp \geq |r_\||$, where $r_\| = (\mathbf{e}_3 \times \mathbf{n}) \cdot \mathbf{r}$ is defined. For $r_\perp \gg a$,

$$F(\mathbf{R}, \mathbf{r}) = \frac{1}{\pi} \left( \frac{a}{\xi_{s\perp}} \right) K_0(r_\perp/\xi_{s\perp}) \cos(k_{F\perp} r_\perp) \delta_{r_\| 0}, \quad (7)$$

where $K_0$ is the modified Bessel function of the zeroth order, $\xi_{s\perp} = a(W/\Delta_0 \sqrt{2})$ is the coherence length, and $k_{F\perp} \sim \pi/a\sqrt{2}$. In the direction perpendicular to the Fermi surface, the extent of the pairing correlations is determined by the coherence length $\xi_{s\perp}$, whereas, in the parallel direction, their range is vanishingly short, $\xi_{s\|} \sim 0$. (ii) For a strong repulsive on-site interaction and a nearest-neighbor attraction, pairing with $d$-wave symmetry may result [19]: $F(\mathbf{R}, \mathbf{r} = 0) = 0$ and $\Delta_\mathbf{k} = 2\Delta_d(\cos k_x a - \cos k_y a)$. Close to half filling and for $r_\perp \gg a$, one can again estimate [20]

$$F(\mathbf{R}, \mathbf{r}) = \frac{\eta}{4\pi} \left( \frac{a}{\xi_\perp} \right) \frac{(r_\|/a)}{[(r_\|/a)^2 + (r_\perp/\xi_\perp)^2]^{3/2}} \sin(k_{F\perp} r_\perp), \quad (8)$$

where $\xi_\perp = a(W/4\Delta_d)$ is the characteristic coherence length perpendicular to the Fermi surface, $\Delta_0 \simeq 4\Delta_d$ is the maximum energy gap at the Fermi surface, and $\eta = \text{sgn}(r_x r_y)$. This result is most useful when there is a clear separation between the length scales $a$ and $\xi_\perp$. For a given $r_\perp$, its maximum value and the position scale as $a\xi_\perp/r_\perp^2$ and $r_\| = a(r_\perp/\xi_\perp\sqrt{2})$. The existence of nodes in the energy gap has resulted in the power-law decay of the pairing correlations in all directions at large distances instead of the exponential one as found in the $s$-wave case. $F$ has $d_{x^2-y^2}$ symmetry as expected.

The interaction between the conduction electrons in the superconductor and the impurity spin is given by the Hamiltonian

$$H_\text{imp} = J\mathbf{S} \cdot \mathbf{s}(0) + Un(0), \quad (9)$$
$$= \frac{1}{N} \sum_{\mathbf{kk}'} \Psi_\mathbf{k}^\dagger \hat{V} \Psi_{\mathbf{k}'},$$

where $\hat{V} = w\hat{\tau}_0 + U\hat{\tau}_3$, $J$ is the exchange coupling, $U$ is the local Coulomb interaction, and the operators $\mathbf{s}(\mathbf{r}) = \frac{1}{2}\sum_{\sigma\nu} \psi_\sigma^\dagger(\mathbf{r})\hat{\tau}_{\sigma\nu}\psi_\nu(\mathbf{r})$ and $n(\mathbf{r}) = \sum_\sigma \psi_\sigma^\dagger(\mathbf{r})\psi_\sigma(\mathbf{r})$ are the conduction electron spin and number densities. Because of spin-rotational symmetry, the coordinate system for the spin degrees of freedom can always be chosen so that the $z$-axis points in the direction of $\mathbf{w} \equiv \frac{1}{2}J\mathbf{S}$. Obviously, the exchange interaction preserves particle-hole symmetry, whereas the Coulomb interaction breaks it.

We will argue, after comparing the uniform order-parameter solution with the self-consistently determined one, that the difference between these two approaches is important only close to the impurity site. The assumption that the spatial variation of the superconducting order parameter can be neglected leads to a considerable simplification of the problem, because the impurity interaction (9) is limited to one site. Scattering of quasiparticles from the impurity moment is described by a $T$ matrix, $\hat{T}(\mathbf{r}, \mathbf{r}'; \omega) = \hat{T}(\omega)\delta_{\mathbf{r}0}\delta_{\mathbf{r}'0}$, which is local and whose Fourier transform therefore is independent of wave vectors. In the Gor'kov-Nambu presentation, the single-particle Green's function is $\hat{G}(\mathbf{r}, \mathbf{r}'; t - t') = -i\langle T\Psi(\mathbf{r}, t)\Psi^\dagger(\mathbf{r}', t')\rangle$. In the presence of an impurity,

$$\hat{G}(\mathbf{r}, \mathbf{r}'; \omega) = \hat{G}^{(0)}(\mathbf{r} - \mathbf{r}', \omega) + \hat{G}^{(0)}(\mathbf{r}, \omega)\hat{T}(\omega)\hat{G}^{(0)}(-\mathbf{r}', \omega), \quad (10)$$

where both $\hat{G}^{(0)}(\mathbf{r}, \omega) = N^{-1}\sum_\mathbf{k} \hat{G}^{(0)}(\mathbf{k}, \omega)e^{i\mathbf{k}\cdot\mathbf{r}}$ and $\hat{T}(\omega)$ are matrices in two dimensional Gor'kov-Nambu space. The single-particle Green's function of a clean system is $\hat{G}^{(0)}(\mathbf{k}, \omega) = (\omega\hat{\tau}_0 - \epsilon_\mathbf{k}\hat{\tau}_3 + \Delta_\mathbf{k}\hat{\tau}_1 + i0^+)^{-1}$. The $T$ matrix can be obtained from the Lippmann-Schwinger equation in the form of

$$\hat{T}(\omega) = [\hat{V}^{-1} - \hat{G}^{(0)}(\mathbf{r} = 0, \omega)]^{-1}. \quad (11)$$

These equations allow a complete solution of the problem as long as order-parameter relaxation can be neglected.

### III. MAGNETIC MOMENT IN AN S-WAVE SUPERCONDUCTOR

Conventional pairing of $s$-wave symmetry is described by the gap function $\Delta(\mathbf{r}) \equiv \Delta(\mathbf{r}, 0)$. It may be modeled by a system with an attractive on-site interaction $v(\mathbf{r}) = v(0)\delta_{\mathbf{r}0}$, $v(0) < 0$.

#### A. Uniform order parameter

First, consider a superconductor at half filling with a spatially uniform $s$-wave gap function, $\Delta(\mathbf{r}) = \Delta_0$. As shown by Yu [1], Shiba [2], and Rusinov [3], the magnetic impurity gives rise to a bound state inside the superconducting energy gap. Their solution suggests that there exists a singular point, $w = w_c(U)$, that separates two distinct ground states of the system. The nature of the bound state can be found by computing the spectral density per site

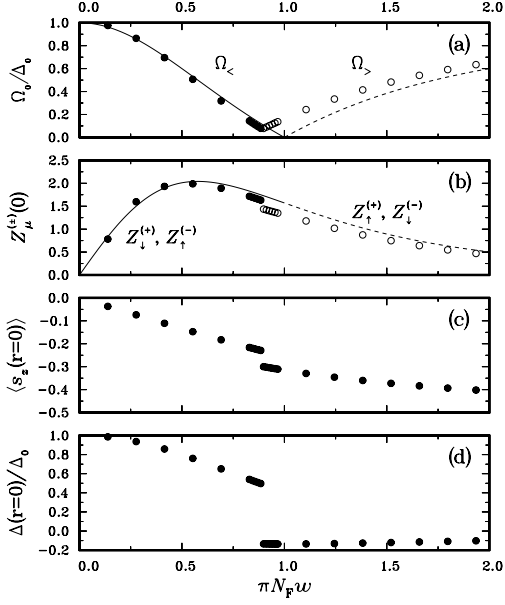

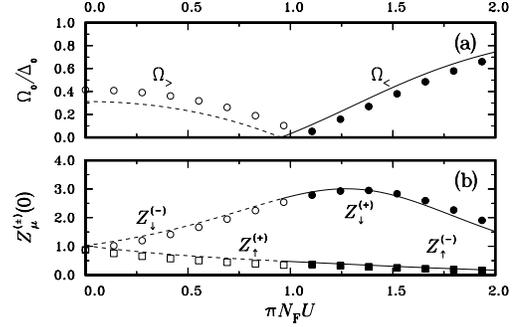

FIG. 3. (a) The bound-state energy $\Omega_0$, (b) the spectral weight $Z_\sigma^{(\pm)}(0)$ (in units of $N_F \Delta_0$), (c) the spin polarization $\langle s_z(\mathbf{r}=0) \rangle$ ($\langle n(\mathbf{r}=0) \rangle = 1$), and (d) the gap function $\Delta(\mathbf{r}=0)$ at the impurity site $\mathbf{r}=0$ as a function of $w$ in the $s$-wave superconductor for $U=0$. Lines denote the $T$ matrix results for the uniform order parameter and symbols the self-consistent mean-field results on a square lattice at half filling. The quantities of the impurity-induced intragap quasiparticle state belonging to the branch $w < w_c$ are denoted by solid lines and solid symbols, whereas those ones belonging to the branch $w > w_c$ are marked by dashed lines and open symbols. Note that $Z_\sigma^{(-)}(0) = Z_{-\sigma}^{(+)}(0)$ because of particle-hole symmetry.

$$A_\sigma(\mathbf{r}, \omega) = -\frac{1}{\pi} \text{Im} \, G_{\sigma\sigma}(\mathbf{r}, \mathbf{r}; \omega). \tag{12}$$

For $|\omega| < \Delta_0$, this can be expressed as $A_\sigma(\mathbf{r}, \omega) = Z_\sigma^{(+)}(\mathbf{r}) \delta(\omega - \Omega_0) + Z_\sigma^{(-)}(\mathbf{r}) \delta(\omega + \Omega_0)$, where $Z_\sigma^{(\pm)}(\mathbf{r})$ describes the particle (hole) weight at positive (negative) energy with spin orientation $\sigma$ ($=\uparrow, \downarrow$) and $\Omega_0$ is the energy of the bound state. Note that $Z_\sigma^{(-)}(\mathbf{r}) = Z_{-\sigma}^{(+)}(\mathbf{r})$ whenever the normal-state energy spectrum $\epsilon_\mathbf{k}$ has particle-hole symmetry and $U=0$ [21].

At half filling, $\mu = 0$, one can show that, for $|\omega| < \Delta_0$, $\hat{G}^{(0)}(\mathbf{r}=0, \omega) = -\pi N(\omega)(\omega \hat{\tau}_0 - \Delta_0 \hat{\tau}_1)/\sqrt{\Delta_0^2 - \omega^2}$. For $\Delta_0 \ll W$, we obtain $N(\omega) = (2/\pi^2 W) \log(4W/\sqrt{\Delta_0^2 - \omega^2})$. In the following, we will approximate $N(\omega) \simeq N(0)$. The logarithmic dependence guarantees that this assumption is reasonable as long as $|\omega|$ is not too close to $\Delta_0$. Note that, for $|\omega| \ll W$, $N(0)$ equals to the normal-state density of states averaged over the energy interval $\pm \Delta_0$ around $\mu = 0$. Thus, it is convenient to define the density of states per site in the normal state at the Fermi energy as $N_F \equiv N(0)$. We can immediately solve the above equations for the bound-state energy and the spectral densities. (i) For $w < w_c(U) = \sqrt{(\pi N_F)^{-2} + U^2}$, the bound-state energy $\Omega_0 = \Omega_<$, where

$$\Omega_< = \eta \Delta_0 \frac{c_+ c_- - 1}{\sqrt{(c_+^2 + 1)(c_-^2 + 1)}}, \tag{13}$$

FIG. 4. (a) The bound-state energy $\Omega_0$ and (b) the spectral weight $Z_\sigma^{(\pm)}(0)$ (in units of $N_F \Delta_0$) of the impurity-induced intragap quasiparticle state as a function of $U$ in the $s$-wave superconductor for $\pi N_F w = 1.4$. Lines denote the $T$ matrix results for the uniform order parameter and symbols the self-consistent mean-field results on a square lattice at half filling. The quantities belonging to the branch $w < w_c(U)$ are denoted by the solid lines and solid symbols, whereas those ones belonging to the branch $w > w_c(U)$ are marked by dashed lines and open symbols. Note that $Z_\sigma^{(-)}(0) \ne Z_{-\sigma}^{(+)}(0)$, because particle-hole symmetry is broken by the on-site Coulomb interaction $U$. We emphasize that while the spectral weights of the intragap peaks are asymmetric their energies relative to the chemical potential are not.

$c_\pm = c_w \mp c_u$, and $\eta = \text{sgn}(c_+ c_-)$. Here, the dimensionless parameters are $c_w = (w/\pi N_F)/(w^2 - U^2)$ and $c_u = (U/\pi N_F)/(w^2 - U^2)$. A straightforward calculation shows that $Z_\downarrow^{(+)}(0) = \bar{Z} + \delta Z$, $Z_\uparrow^{(-)}(0) = \bar{Z} - \delta Z$, and $Z_\uparrow^{(+)}(0) = Z_\downarrow^{(-)}(0) = 0$, with $\bar{Z} = (N_F \Delta_0) 2\pi |c_w|[(c_w^2 + c_u^2) + (c_w^2 - c_u^2)^2]/[(c_+^2 + 1)(c_-^2 + 1)]^{3/2}$ and $\delta Z = (N_F \Delta_0) \text{sgn}(U) 4\pi |c_u| c_w^2 / [(c_+^2 + 1)(c_-^2 + 1)]^{3/2}$. Hence, the bound state is a local quasiparticle state with spin down [22]; i.e., the spin projection of the particle component is antiparallel and that of the hole component is parallel to the impurity moment $\mathbf{w}$. Basic features of the low-energy spectral density are illustrated in Fig. 1. As $w$ increases, the minimum energy to make a quasiparticle excitation from the condensate decreases, until, for $w > w_c(U)$, it would be negative. At this point, the superconducting condensate becomes thermodynamically unstable against spontaneous creation of a local quasiparticle excitation with spin down. As a result, the superconductor goes through a quantum transition from the spin-unpolarized to the spin-polarized state. Note that the average number of electrons remains the same. While, for $U=0$, $\langle n(\mathbf{r}) \rangle$ is a uniform function, $\langle s_z(\mathbf{r}) \rangle$ becomes more peaked at $\mathbf{r} = 0$ as $w$ increases. (For $U \neq 0$, also $\langle n(\mathbf{r}) \rangle$ would be nonuniform.) However, the true nature of the transition is revealed by considering the total spin polarization, $\langle s_z \rangle = \sum_{\mathbf{r}} \langle s_z(\mathbf{r}) \rangle$. At zero temperature, we obtain $\langle s_z \rangle = 0$, for $w < w_c(U)$, and $\langle s_z \rangle = -\frac{1}{2}$, for $w > w_c(U)$. Because the ground state has qualitatively changed, so have the elementary excitations, and, in particular, the nature of the impurity-induced bound state. (ii) For $w > w_c(U)$, the energy of the bound state is given by $\Omega_0 = \Omega_>$, with $\Omega_> = -\Omega_<$. Moreover, it is now associated with a quasiparticle with spin parallel to the impurity moment: $Z_\uparrow^{(+)}(0) = \bar{Z} - \delta Z$, $Z_\downarrow^{(-)}(0) = \bar{Z} + \delta Z$ and $Z_\downarrow^{(+)}(0) = Z_\uparrow^{(-)}(0) = 0$. The bound state in both cases accounts for one electronic degree of freedom.

Figure 3 summarizes the above results for $U = 0$ and Figure 4 describes the effect of nonzero $U$. The Coulomb interaction effectively increases the critical coupling $w_c(U)$. Thus, for a large enough $U$, the doublet state is thermodynamically stable. It is also clear that the on-site Coulomb interaction breaks up the particle-hole symmetry. For example, given that the impurity moment can be described classically and the envelop of the experimentally observed intragap spectral weight of the electronlike state $Z^{(+)}(\mathbf{r})$ is larger than that of the holelike state $Z^{(-)}(\mathbf{r})$, then either (a) $U > 0$ and $w < w_c(U)$ or (b) $U < 0$ and $w > w_c(U)$.

Interestingly, one is tempted to interpret the recent experiment [11] on magnetic impurities as if the intragap spectral density decays on a length scale determined not by the coherence length but by a much shorter scale of the order of the Fermi length $\ell_F$. This result appears paradoxical because a simple scaling estimate suggests that if the extent of the bound state were $\mathcal{O}(\ell_F)$, the energy of this state should be $\hbar v_F/\ell_F \sim \epsilon_F$ in a one-band picture and therefore could not be localized within the energy gap. One possible explanation of the experimental result is the power-law prefactor of the exponentially-decaying bound state. For example, consider a uniform gap function, $\Delta(\mathbf{r}) = \Delta_0$, and an isotropic energy band. In three dimensions, we find that $Z_\sigma^{(\pm)}(\mathbf{r}) \propto (\ell_F/r)^2 e^{-2r/\lambda_0}$, where $\lambda_0 = \xi_0/\sqrt{1 - (\Omega_0/\Delta_0)^2}$, $\xi_0 = \hbar v_F/\Delta_0$, and $\ell_F = 2\pi/k_F$. Thus, at the distance $r = 10\ell_F$, $Z_\sigma^{(\pm)}(\mathbf{r})$ has already decreased by a factor of $\sim 100$ from its maximum value, which certainly should be enough to explain the experimental findings. Indeed, the spatial variation of the order parameter in the neighborhood of the impurity does not seem to play a major role in determining the behavior of $Z_\sigma^{(\pm)}(\mathbf{r})$, even though the order parameter relaxes within a few Fermi lengths $\ell_F$ [24].

### B. Non-uniform order parameter

A qualitative understanding of the consequences of the order-parameter relaxation is obtained by considering a local suppression of the gap function at the impurity site, $\delta\Delta = \Delta_0 - \Delta(\mathbf{r} = 0)$. Its effect can be accounted for by including the additional term $\delta\Delta\Psi^\dagger(0)\hat{\tau}_1\Psi(0)$ in Hamiltonian (9). Thus, $\hat{V} = w\hat{\tau}_0 + U\hat{\tau}_3 + \delta\Delta\hat{\tau}_1$. First, consider $w = 0$. The suppression of the order parameter provides a local attractive potential for quasiparticles which produces two bound states in the superconducting energy gap at the energy

$$\Omega_0 = \Delta_0 \sqrt{1 - \alpha^2}, \qquad (14)$$

where $\alpha = c_\Delta/\beta$. In addition, we have defined $\beta^2 = \frac{1}{4}(1 - c^2)^2 + c_w^2$ ($\beta \geq 0$), $c^2 = c_w^2 - c_u^2 - c_\Delta^2$, $c_w = (w/\pi N_F)/(w^2 - U^2 - \delta\Delta^2)$, $c_u = (U/\pi N_F)/(w^2 - U^2 - \delta\Delta^2)$, and $c_\Delta = (\delta\Delta/\pi N_F)/(w^2 - U^2 - \delta\Delta^2)$. In contrast to the bound state induced by the magnetic moment, this state is two-fold degenerate by virtue of time-reversal symmetry. Their spin-quantum numbers are $\pm\frac{1}{2}$. Note that the bound states become essentially indistinguishable from the quasiparticle continuum (i.e., $\alpha^2 \ll 1$) in the limits of the weak order-parameter suppression, $\pi N_F \delta\Delta \ll 1$, and the strong potential scatterer, $\pi N_F |U| \gg 1$. Positive $w$ breaks the degeneracy by lowering the energy of the spin-down state and increasing the energy of the spin-up state:

$$\Omega_0/\Delta_0 = \frac{1 - c^2}{2\beta}\sqrt{1 - \alpha^2} \pm \frac{c_\Delta c_w}{\beta^2}. \qquad (15)$$

For $w \gg \delta\Delta$, the spectral density measured at the impurity site $\mathbf{r} = 0$ is much larger for the former state at the energy $\omega \lesssim \Delta_0$ than for the latter one whose spectral density is either transformed into the quasiparticle continuum or to the antibound state above the energy $W$.

Next, compare the nonself-consistent $T$ matrix predictions with the results obtained when the gap function is allowed to relax in the neighborhood of the impurity moment. Our numerical approach is based on the model defined by Eqs. (3) and (9) on a square lattice with the effective on-site electron-electron attraction of magnitude $v(0)$ such that $\Delta_0/W = 0.05$. The model, together with the mean-field equation, $\Delta(\mathbf{r}) = -v(0)F(\mathbf{r}, 0)$, is solved self-consistently (e.g., in the Bogoliubov – de Gennes scheme) [23]. As expected, the overall agreement between it and the $T$ matrix calculation is good; see Figs. 3 and 4. However, the additional variational degree of freedom shifts $w_c$ to a lower value, and $\Omega_0$ never becomes exactly zero. For $w > w_c$, $\Delta(\mathbf{r} = 0)$ is overscreened to a negative value and, ultimately, $\Delta(\mathbf{r} = 0) \to 0$, as $w \to \infty$. The non-zero total spin polarization residing in the neighborhood of the impurity has lead to a nearly complete pair-breaking effect at this site. The charge density remains uniform everywhere in the system for arbitrary $w$ and $U = 0$. For $w \sim w_c(U)$, the results that assume a uniform gap function everywhere deviate most clearly from those ones allowing the gap function relax locally, because $\delta\Delta/W$ and $\delta\Delta/w$ have their largest values for the same values of $w$.

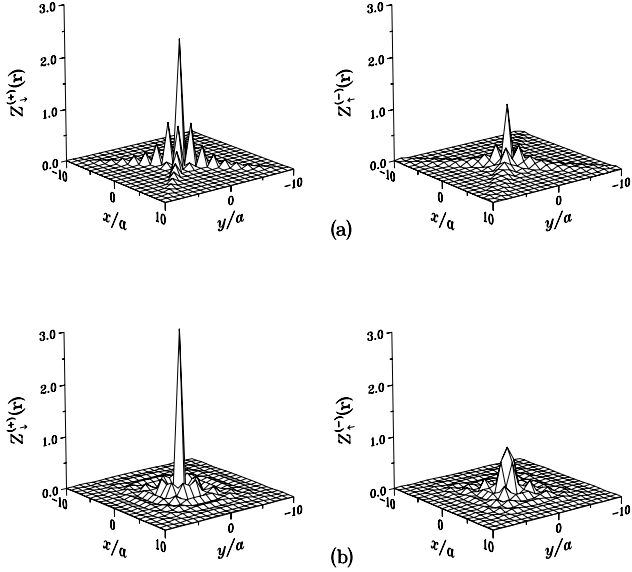

FIG. 5. The spectral weight $Z_\downarrow^{(+)}(\mathbf{r})$ and $Z_\uparrow^{(-)}(\mathbf{r})$ (in units of $N_F\Delta_0$) of the impurity-induced intragap quasiparticle state in the $s$-wave superconductor for (a) $\mu = 0$, $\pi N_F U = 0.4$, and (b) $\mu = -W/2$, $U = 0$. These results are computed self-consistently on a square lattice with the lattice spacing $a$, $\pi N_F w = 0.7$, and $\Delta_0/W = 0.05$. The bound-state energies are (a) $\Omega_0 = 0.39\Delta_0$ and (b) $\Omega_0 = 0.46\Delta_0$.

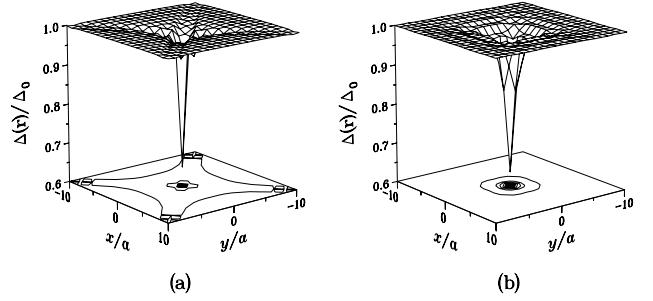

FIG. 6. The self-consistent $s$-wave gap function $\Delta(\mathbf{r})$ around a magnetic moment for (a) $\mu = 0$, $\pi N_F U = 0.4$, and (b) $\mu = -W/2$, $U = 0$, obtained on a square lattice with the lattice spacing $a$. The other parameters are $\pi N_F w = 0.7$ and $\Delta_0/W = 0.05$.

At $\mu = 0$, the Fermi surface is a square producing a very anisotropic Fermi velocity. This anisotropy causes the impurity-induced bound state to have a shape which extends along the diagonal directions; see Fig. 5(a). A similar spatial form has also been observed in the context of relaxed quasiparticle excitations [25]. Notably, the decay of the spectral density due to the bound state is governed by $\xi_\mathbf{n} = \hbar v_\mathbf{n}/\Delta_0$, where $\mathbf{n}$ is a normal to the Fermi surface at a given point and $v_\mathbf{n} = \hbar^{-1}(\mathbf{n} \cdot \nabla_\mathbf{k}\epsilon_\mathbf{k})_{k=k_F}$ is the Fermi velocity at the same point. For a flat Fermi surface, $Z_\sigma^{(\pm)}(\mathbf{r}) \propto \delta_{r_\parallel 0} e^{-2r_\perp/\lambda_\perp}$, where $r_\parallel$ and $r_\perp$ are the components of $\mathbf{r}$ parallel and perpendicular to the sections of the Fermi surface for which $r_\perp \geq |r_\parallel|$. Here, $\lambda_\perp = \xi_{s\perp}\sqrt{1-(\Omega_0/\Delta_0)^2}$. This is in contrast to the isotropic $r^{-1}e^{-2r/\lambda_0}$ behavior found in two dimensions for a circular Fermi surface. The anisotropy in the Fermi velocity yields a similar structure in the gap function around the magnetic impurity [26]. The slower relaxation rate along the diagonal directions is clearly observable in Fig. 6(a).

Moving away from half filling, the anisotropy in the Fermi velocity decreases, and the spatial form of the impurity-induced bound state and the gap function in the neighborhood of the impurity become increasingly isotropic, as shown by Figs. 5(b) and 6(b). Moreover, for $\mu \neq 0$, the envelopes of the spectral weights $Z_\downarrow^{(+)}(\mathbf{r})$ and $Z_\uparrow^{(-)}(\mathbf{r})$ differ, because the single-particle band structure of the normal state is no longer particle-hole symmetric. The sensitivity to the energy spectrum away from the vicinity of the chemical potential is a consequence of strong impurity scattering whose spatial extent is less than the lattice spacing and which efficiently mixes quasiparticle states with a wide wave-vector range around $k_F$. A similar conclusion is reached in Ref. [14] where the effect of the band structure is discussed.

The deformation of the order parameter in the vicinity of the magnetic moment creates an attractive potential with a finite range. This feature may lead to additional bound states which are no longer invariant under all symmetry operations of the square lattice (i.e., to higher "angular momentum" states) but whose binding energies are very small. Because these states have a node at the impurity site, they do not couple to the impurity moment and they are at least two-fold degenerate due to the spin degeneracy. In general, we find a two-fold degenerate state slightly below $\Delta_0$ that transforms like quadrupole ($x^2 - y^2$). However, away from half filling and for $w < w_c$, there is some evidence for the lowest energy, nonsymmetric quasiparticle state that is four-fold degenerate and transforms like dipole.

## IV. MAGNETIC MOMENT IN A $D$-WAVE SUPERCONDUCTOR

Invariance of various perturbations under particle-hole transformation was significant in classifying the properties of the impurity-induced quasiparticle states in $s$-wave superconductors. In a similar fashion, we may de-scribe the same properties in $d$-wave superconductors under particle-hole transformation. In $d$-wave superconductors, it is useful to introduce still another transformation, namely "charge conjugation": $\Psi_\mathbf{k} \to \Psi'_\mathbf{k} = \hat{\tau}_2 \Psi_{\mathbf{Q}-\mathbf{k}}$. It identifies the cases when the spin-unpolarized ground state of the superconductor interacting with a magnetic moment is unstable against spontaneous creation of a local spin polarization. As defined here, an $s$-wave superconductor is not invariant under charge conjugation.

### A. Invariance under charge conjugation

First, we consider a quasi-two-dimensional system whose energy spectrum is particle-hole symmetric in the normal state and whose superconducting energy gap has $d_{x^2-y^2}$ symmetry at the Fermi energy: $\Delta_\mathbf{k} = \Delta_0 \cos 2\varphi$. We also assume that the gap function is uniform everywhere and that the Fermi surface in the normal state is circular. Since the system possesses charge-conjugation symmetry, $\hat{\tau}_2 \hat{G}^{(0)}(\mathbf{r}=0,\omega)\hat{\tau}_2 = \hat{G}^{(0)}(\mathbf{r}=0,\omega)$, it is immediately clear that $\hat{G}^{(0)}(\mathbf{r}=0,\omega) = G_0(\omega)\hat{\tau}_0$. At the impurity site, the spectral density has a particularly simple form:

$$A_\sigma(\mathbf{r}=0,\omega) = -\frac{1}{\pi}\text{Im}\frac{G_0(\omega)}{1-\sigma v_\sigma G_0(\omega)}, \quad (16)$$

where $v_\pm = w \pm U$, $\sigma = +$ for spin up, and $\sigma = -$ for spin down. The poles of $A_\sigma(\mathbf{r}=0,\omega)$ give information about the quasiparticle states. At low energies $|\omega| \ll \Delta_0$, one obtains the formula [5,6] $G_0(\omega) = -N_F(2\omega/\Delta_0)[\log(4\Delta_0/|\omega|) + i\,\text{sgn}(\omega)\pi/2]$. For $|c_\pm| \ll 1$, where $c_\pm = (\pi N_F v_\pm)^{-1}$, the magnetic moment induces two quasiparticle states at energies $\Omega_\pm = \frac{\pi}{2} c_\pm \Delta_0 / \log(8/\pi|c_\pm|)$ with inverse lifetimes $\Gamma_\pm = \frac{\pi}{2}|\Omega_\pm|/\log(8/\pi|c_\pm|)$. For $U = 0$, the two states are degenerate: $\Omega_+ = \Omega_-$. Their spin quantum numbers are $s_z = -\frac{1}{2}$. With increasing $U$, the degeneracy is lifted as $\Omega_+$ decreases and $\Omega_-$ increases. For $U \ll w$, they appear as resonances in the spectral densities $A_\uparrow(\mathbf{r}=0,\omega)$ and $A_\downarrow(\mathbf{r}=0,\omega)$ at $\omega = -\Omega_+$ (holelike) and $\omega = \Omega_-$ (electronlike), respectively. For $U \sim w$, there is no solution for $\Omega_-$ and no resonance structure develops in $A_\downarrow(\mathbf{r}=0,\omega)$. For $U \gg w$, $\Omega_-$ is negative and approaches zero from below. Thus, for each spin orientation, the spectral density $A_\sigma(\mathbf{r}=0,\omega)$ has a holelike resonance at the energy $\omega = -\Omega_\sigma$. However, the quasiparticle energies $\Omega_\pm$ never cross the chemical potential for any value of $w$ and $U$. This ensures that the superconducting ground state remains stable against spontaneous creation of quasiparticle excitations. Thus, in contrast to the $s$-wave case, a superconductor with $d$-wave pairing does not exhibit a transition where the spin-quantum number of the ground state changes from zero to $\frac{1}{2}$. The difference follows from charge-conjugation symmetry which leads to vanishing $\hat{G}^{(0)}(\mathbf{r}=0,\omega)$ at the chemical potential. This prevents $\Omega_\pm$ for crossing the chemical potential and ultimately pins it to zero energy as $w \to \infty$ with $U$ fixed.

In the neighborhood of the impurity, the virtual-bound state acquires the characteristic form of the $d$-wave energy gap with tails extending along the diagonal directions. The spatial form of this state can be obtained by computing the spectral density, $A_\sigma(\mathbf{r},\omega) = A_0(\omega) + \delta A_\sigma(\mathbf{r},\omega)$, where $A_0(\omega) = -\pi^{-1}\text{Im}\,G_0(\omega)$ is the uniform spectral density of a clean superconductor and $\delta A_\sigma(\mathbf{r},\omega)$ is the impurity-induced contribution. For $|\omega| \ll \Delta_0$, $A_0(\omega) = N_F|\omega|/\Delta_0$. The impurity-induced contribution is

$$\delta A_\sigma(\mathbf{r},\omega) = -\frac{1}{\pi}\text{Im}[G_+^{(0)}(\mathbf{r},\omega)T_+^{(\sigma)}(\omega)G_+^{(0)}(-\mathbf{r},\omega) \\ + G_1^{(0)}(\mathbf{r},\omega)T_-^{(\sigma)}(\omega)G_1^{(0)}(-\mathbf{r},\omega)], \quad (17)$$

where $T_\pm^{(\sigma)}(\omega) = -1/[G_\pm(\omega) - \sigma v_\pm^{-1}]$. We use the notation in which $G_\alpha^{(0)}(\mathbf{r},\omega) = \frac{1}{2}\text{Tr}\,\hat{\tau}_\alpha \hat{G}^{(0)}(\mathbf{r},\omega)$ and $G_\alpha(\omega) = G_\alpha^{(0)}(\mathbf{r}=0,\omega)$ ($\alpha = 0,\ldots,3$). In addition, we have defined $G_\pm^{(0)}(\mathbf{r},\omega) = G_0^{(0)}(\mathbf{r},\omega) \pm G_3^{(0)}(\mathbf{r},\omega)$ and $G_\pm(\omega) = G_\pm^{(0)}(\mathbf{r}=0,\omega)$. Note that particle-hole symmetry implies the relation $G_\pm(\omega) = G_0(\omega)$. For $U = 0$, $T_+^{(\sigma)}(\omega) = T_-^{(\sigma)}(\omega)$, and the spectral density is particle-hole symmetric: $\delta A_\uparrow(\mathbf{r},-\omega) = \delta A_\downarrow(\mathbf{r},\omega)$. While the overall spatial dependence of the virtual-bound states generated by purely magnetic ($U=0$) and nonmagnetic ($w=0$) scattering is the same, the resonance factors $T_\pm^{(\sigma)}$ lead to distinct states. The specific form of the spectral density is determined by the Green's function, $\hat{G}^{(0)}(\mathbf{r},\omega)$. Eq. (17) produces a qualitatively correct description of the quasiparticle states as long as the energy gap is not comparable to the bandwidth, because then the relaxation of the gap function in the vicinity of the defect can be neglected as a small correction. Nonetheless, the local suppression of the gap function $\Delta(\mathbf{r}=0)$ could easily be included. For $\omega = \mp\Omega_\pm$, the spectral density is most strongly enhanced in the horizontal and vertical directions (the extrema directions of the $d$-wave energy gap) at distances $r \ll \xi_0$ ($= \hbar v_F/\Delta_0$) from the impurity. Further away from the impurity ($r \sim \xi_0$), the situation is reversed: the spectral density is enhanced in the neighborhood of the diagonal directions at the resonances.

That the Green's function $\hat{G}^{(0)}(\mathbf{r}=0,\omega)$ is proportional to $\hat{\tau}_0$ has curious consequences which set $d$-wave pairing apart from $s$-wave one. In the $s$-wave superconductor, low-energy quasiparticles have both particle and hole character due to the non-zero pairing field $F(\mathbf{R},\mathbf{r}=0)$. In contrast, $d$-wave symmetry forces the pairing field to vanish locally, $F(\mathbf{R},\mathbf{r}=0) = 0$: the particle and hole degrees of freedom are effectively decoupled at the impurity site. Strong magnetic scattering leads to two virtual bound states whose components at the impu-rity site are either electronlike with spin down or holelike with spin up; the other components vanish because their orbital character is the same as that of the order parameter (i.e., they have a node at the impurity site). A nonzero $U$ lifts the degeneracy without mixing the states. Note that, for an impurity whose non-magnetic scattering strength ($U > 0$) is larger than the magnetic one, two virtual-bound states of hole character with quantum numbers $s_z = \pm\frac{1}{2}$ are obtained at the impurity site [6].

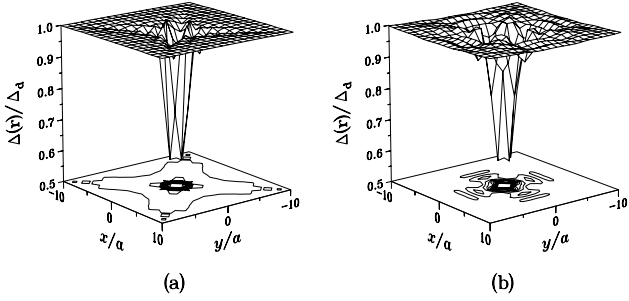

FIG. 7. The self-consistent $d$-wave gap function [28] $\Delta(\mathbf{r})$ around a magnetic moment ($\pi N_F w = 10$ and $U = 0$) for (a) $\mu = 0$ and (b) $\mu = -W/2$ obtained on a square lattice with the lattice spacing $a$ and the coherence length $\xi_\perp = 10a$ ($4\Delta_d/W = 0.1$). The minimum at $\mathbf{r} = 0$ is not shown. $\Delta_d$ is the spatially uniform $d$-wave gap function in the clean system. Note that the maximum energy gap at the Fermi surface is (a) $\Delta_0 = 4\Delta_d$ and (b) $\Delta_0 = 2\Delta_d$.

In conclusion, a $d$-wave superconductor appears locally as a simple metal with a pseudogap: at short distances, the particle and hole excitations are decoupled and the energies of the virtual-bound states do not cross the chemical potential due to the level repulsion. Another treatment in terms of elementary quasiparticle excitations is outlined in Appendix A.

### B. Broken charge-conjugation symmetry

First, assume a mixed $s$- and $d$-wave superconductor so that the pairing state is described by a gap function $\Delta_{\mathbf{k}} = \Delta_0 \cos 2\varphi + \Delta_s$, where a small $s$-wave component has been added to the $d$-wave state, $\Delta_s \ll \Delta_0$. This type of admixture of two angular-momentum states is relevant in a $d$-wave superconductor where the tetragonal symmetry is broken by an orthorhombic distortion. It also breaks charge-conjugation symmetry. Under these conditions, $\hat{G}^{(0)}(\mathbf{r} = 0, \omega) = G_0(\omega)\hat{\tau}_0 + G_1(\omega)\hat{\tau}_1$; for $|\omega| \ll \Delta_0$, $G_0(\omega)$ is given above and $G_1(\omega) \simeq \pi N_F \Delta_s/\Delta_0$. For purely magnetic scattering ($U = 0$), the energies of the virtual-bound states, $\Omega_\pm = \frac{\pi}{2}\tilde{c}_\pm \Delta_0/\log(8/\pi|\tilde{c}_\pm|)$, where $\tilde{c}_\pm = c_w \pm \Delta_s/\Delta_0$, are split by the $s$-wave component. Clearly, the state at $\omega = \Omega_-$ crosses the chemical potential for $c_w < c_* \equiv \Delta_s/\Delta_0$. The critical coupling $c_*$ signifies a quantum transition from the spin-unpolarized to a spin-polarized state of the superconductor. We remark that, for an $s$-wave superconductor, $c_* = 1$. As $\Delta_s \to 0$, the critical coupling $c_* \to 0$, and the quantum transition is shifted to higher values of the magnetic moment until, for $\Delta_s = 0$, the transition is completely suppressed.

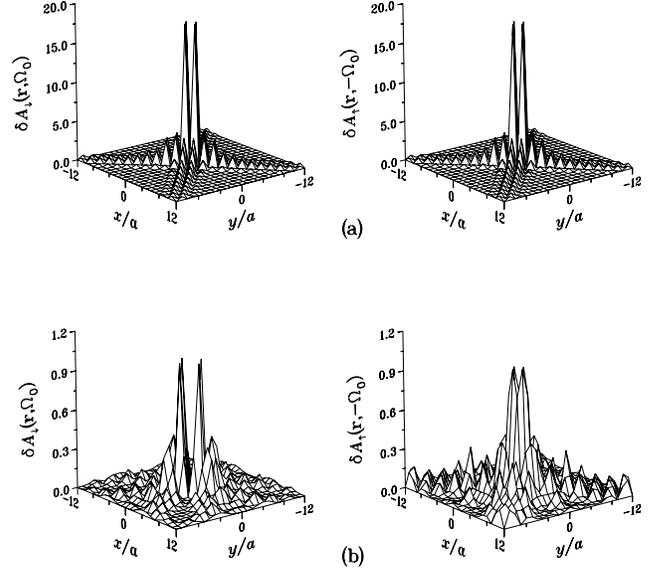

FIG. 8. The spectral density $\delta A_\sigma(\mathbf{r}, \pm\Omega_0)$ (in units of $N_F$) for (a) $\mu = 0$ and (b) $\mu = -W/2$ in a two-dimensional $d$-wave superconductor as a function of position $\mathbf{r} = (x, y)$ around a classical magnetic moment ($\pi N_F w = 10$ and $U = 0$) located at $\mathbf{r} = 0$; $a$ is the lattice spacing. These results are computed self-consistently with $\xi_\perp = 10a$. At half filling, the spectral density obeys particle-hole symmetry: $\delta A_\uparrow(\mathbf{r}, -\Omega_0) = \delta A_\downarrow(\mathbf{r}, \Omega_0)$. The energies of the shown virtual-bound states are (a) $\Omega_0 \simeq 0.05\Delta_0$ and (b) $\Omega_0 \simeq 0.5\Delta_0$.

Second, if the particle-hole symmetry is broken in the normal state, the Green's function acquires a non-zero $\hat{\tau}_3$ component. For a large enough magnetic moment, it will again induce a transition where the spin-quantum number is changed from zero to $\frac{1}{2}$.

In general, both of the above features may be present simultaneously. In such a case, one must replace $\tilde{c}_\pm$ by $c_w \pm \sqrt{g_1^2 + (c_u + g_3)^2}$, where $g_\alpha = (\pi N_F)^{-1} G_\alpha(\Omega)$ ($\alpha = 1, 3$) and $w > |U|$. Similarly, $c_*$ is replaced by $c_w$ for which $\tilde{c}_- = 0$. As a consequence, the critical coupling becomes a function of the chemical potential and the $s$-wave component: $c_* = c_*(\mu, \Delta_s)$.

### C. Non-uniform order parameter

The Fermi-velocity distribution affects the precise form of the $d$-wave gap function [27] in the neighborhood of the magnetic moment; see Fig. 7. As in the $s$-wave superconductor, the sensitivity to the Fermi-surface geometry originates from the mode composition of the gap function: the electronic degrees of freedom within the energy proportional to $\Delta_0$ about the chemical potential have the largest weight in the gap function. Close to half filling, the gap function has four troughs extending along the diagonal directions, whereas, away from half filling, the four-fold symmetry of the energy gap is more crucial in determining the spatial relaxation of the gap function. That the quasiparticle states are concentrated along the horizontal and vertical directions at energies for which the density of states is the highest ($\omega \sim \Delta_0$) explains the latter finding. In the strong-coupling limit, some modifications are expected [29].

A strong magnetic moment creates virtual-bound states whose spatial dependence has distinctive features due to $d$-wave pairing symmetry. For example, in the neighborhood of the defect, both the particle- and hole components of the spectral density remain non-zero along the diagonals, although the maximum value at a given distance is located off-diagonally; see Fig. 8. Close to half filling, the Fermi-surface geometry has a strong focusing effect on the spatial variation of the virtual-bound states, and their lobes along the vertical directions spread only slowly. At half filling ($\mu = 0$), the lattice Green's functions $G_1^{(0)}(\mathbf{r}, \omega)$ and $G_3^{(0)}(\mathbf{r}, \omega)$ vanish identically at the sites for which $(x+y)/a$ is an even integer. Because, for $|\omega| \ll \Delta_0$, $G_0^{(0)}(\mathbf{r}, \omega)$ is very small, the spectral density is negligible along these lines. Note that, in general, $G_1^{(0)}(\mathbf{r}, \omega)$ does not vanish at the same lattice sites for $s$- and $d$-wave pairing, explaining the difference between Figs. 5(a) and 8(a). The spatial variation of the virtual-bound states is accurately given by Eq. (17) even though it assumes a uniform gap function as one can verify in this case. For $r_\perp \gg a$ and $U=0$, one can estimate

$$\delta A_\sigma(r, \omega \sim 0) \propto \sin^2(k_{F\perp} r_\perp)/[(r_\parallel/a)^2 + (r_\perp/\xi_\perp)^2], \quad (18)$$

where $r_\perp \geq r_\parallel$ (by definition). At half filling, $k_{F\perp} r_\perp = \pi(x+y)/2a$. Away from half filling ($\mu = -W/2$), the spectral density tends to vary radially demonstrating more clearly the effect of the $\mathbf{k}$-space structure of the superconducting energy gap. The example shown in Fig. 8(b) is computed in the spin-polarized ground state which is obtained when the magnetic moment is large enough and the superconductor is not half filled. Neither does the spectral density obey particle-hole symmetry in this case. In general, quasiparticle excitations are qualitatively described by Eq. (17) which provides a straightforward method for determining the spectral density and the resonance energies. For example, the resonance energies are obtained from the equation, $G_\pm(\Omega) = 1/v_\pm$. The two-fold degeneracy of the virtual bound state is lifted both by the local Coulomb interaction and by the band structure which is asymmetric relative to the chemical potential. Because the lattice Green's functions are fast to compute numerically, we have been able study in detail the evolution of the spectral density as a function of the chemical potential and the impurity potential (not shown).

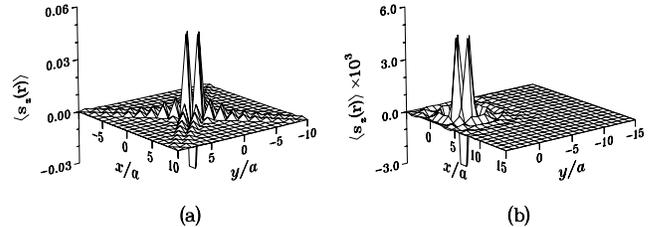

FIG. 9. The electron spin density $\langle s_z(\mathbf{r}) \rangle$ in a two-dimensional $d$-wave superconductor as a function of position $\mathbf{r} = (x, y)$ around a classical magnetic moment ($\pi N_F w = 10$ and $U = 0$) located at $\mathbf{r} = 0$; $a$ is the lattice spacing. The chemical potential is (a) $\mu = 0$ and (b) $\mu = -W/2$. The peak $\langle s_z(\mathbf{r}=0) \rangle \sim -\frac{1}{2}$ in both cases is cut off in order to illustrate the finer details. These results are computed self-consistently with $\xi_\perp = 10a$. Note that (a) $\langle s_z \rangle = 0$ and (b) $\langle s_z \rangle = -\frac{1}{2}$.

In spite of the possibility of the quantum transition, $\Delta(\mathbf{r}=0)$ does not change sign for any value of magnetic moment $w$. While, for $c_* \lesssim c_w \ll 1$, the electron spin density is nearly completely polarized at the impurity site, $\langle s_z(\mathbf{r}=0) \rangle \simeq -\frac{1}{2}$, there is a compensating spin-density cloud of opposite sign in the neighborhood of the magnetic moment that leads to the vanishing spin polarization. Figure 9 shows the spatial variation of the electron spin density with a given magnetic moment both at and away from half filling, illustrating the two cases $c_w > c_*(\mu)$ ($\mu = 0$) and $c_w < c_*(\mu)$ ($\mu \neq 0$). The magnetic moment perturbs electronic degrees of freedom on the energy scale of $w$ which must be large compared to the superconducting energy gap $\Delta_0$ to have pronounced virtual-bound states. At half filling, the behavior of the spin density is dominated by the anisotropy in the Fermi velocity, disguising the effect of pairing symmetry. Away from half filling, the virtual-bound states contribute at distances $r > \xi_0$, while at shorter distances the non-zero spin density is mostly coming from quasiparticle states residing outside the energy-gap region.

These considerations offer intriguing prospects regarding high-temperature superconductors. In addition to possible $d$-wave features, their low carrier densities exacerbate the role of the highly anisotropic Fermi-surface geometry that is inevitably mixed in the spatial dependence of the gap function and the quasiparticle states in the vicinity of impurities. These properties should be accessible to many experimental probes, such as scanning-tunneling microscopy, etc. In particular, it might now be possible to determine experimentally the coherence length in short-coherence-length superconductors by di-rectly measuring the spectral density in the vicinity of defects. In low-temperature superconductors, this is not possible because the power-law prefactors kill the signal before the exponential cutoff set by the coherence length can be observed. Even though the coherence length in these materials is typically short ($\xi_0 \sim 20$Å), it still is long enough that our mean-field results should be relevant in describing their properties qualitatively [30].

## V. MANY MAGNETIC IMPURITIES

As an example, consider strongly-scattering magnetic impurities in an $s$-wave superconductor. In the dilute limit of impurities, only nearest-neighbor impurity states interact due to their exponentially decreasing overlaps. Appendix B describes a mapping of a system of dilute impurities onto an effective model. Using known results for such a model, we conclude that all quasiparticle states in the impurity band in two dimensions are localized. In three dimensions, there is a critical density of impurities, below which all states are localized and above which a mobility edge appears. A similar problem of the impurity-induced quasiparticle states states in a two-dimensional $d$-wave superconductor has also been considered [31].

Although the assumption that the impurities do not interact leads to an equal density of states at every impurity site, the fact that the impurity-induced quasiparticle states may have large overlaps modifies this conclusion. Disorder with strongly interacting impurities yields large impurity- and energy-dependent variations in the quasiparticle wave functions. These variations may, in the worst case, obscure the detection of bound states — for example, by scanning tunneling microscopy — or facilitate the detection if the probe is close to impurity sites that have very large relative weights in the wave function.

## VI. FINAL REMARKS

There is a qualitative difference between $s$- and $d$-wave superconductors regarding the quasiparticle states induced by magnetic moments. This difference arises from the local properties of the superconducting order parameter $F(\mathbf{R},\mathbf{r})$ which is non-zero for $s$-wave pairing but vanishes for $d$-wave pairing at $\mathbf{r} = 0$. Thus, in the $s$-wave superconductor, any local probe sees particle and hole degrees of freedom that are coupled by the pairing field. This feature suggests that the two intragap peaks in the spectral density should be interpreted as reflections of the same quasiparticle state. In contrast, because the pairing field in the $d$-wave superconductor vanishes at short distances, a local perturbation leads to two separate quasiparticle states which are either purely particle- or purely holelike at the impurity site. One can distinguish the two cases by their response to a local Coulomb interaction $Un(0)$. The properties of the quasiparticle excitations are affected by $U$ because their charges are not the same. Using the ground-state charge as a reference point, define the charge operator $Q$ as

$$Q = \sum_{\mathbf{r}}[n(\mathbf{r}) - \langle\Phi_0|n(\mathbf{r})|\Phi_0\rangle]. \qquad (19)$$

Clearly, $\langle\Phi_{-1\downarrow}|Q|\Phi_{-1\downarrow}\rangle = u_1^2 - v_1^2$. If the particle and hole features belong to two distinct states, the Coulomb interaction $U$ will shift their energies differently because their charges differ. In contrast, if they are manifestations of the same state, the Coulomb interaction could not affect them independently, and the spectral density will have two intragap peaks that are symmetrically located below and above the chemical potential. Indeed, we always find that, in an $s$-wave superconductor, the spectral density contains two peaks at energies $\pm\Omega_0$ *regardless of the local Coulomb interaction* whereas, in a $d$-wave superconductor, the Coulomb interaction shifts the quasiparticle peaks asymmetrically relative to the chemical potential.

We conclude by noting that the special nature of the induced quasiparticle states has interesting implications for optical absorption in an $s$-wave superconductor containing a local moment. Because the bound state inside the energy gap exists only for a given spin direction (its time-reversed conjugate overlaps with the quasiparticle continuum), no sharp absorption feature is found at $2\Omega_0$ but the optical absorption begins at the energy $\Delta_0 + \Omega_0$.

## ACKNOWLEDGMENTS

We would like to thank J. Byers and M. Flatte for interesting discussions and, in particular, for helpful comments on the Fridel oscillations in the normal state, A. Yazdani for discussions on experimental issues, C. Kallin for introducing M.I.S. to the Bogoliubov – de Gennes technique, and A.J. Berlinsky for useful comments on the manuscript. This work was supported in part by Natural Sciences and Engineering Research Council of Canada, the Ontario Center for Materials Research (M.I.S.), by the U.S. Department of Energy (A.V.B.), by the NSF Grant No. DMR-9629987 (J.R.S.), and by the Many Body Theory Program at Los Alamos.

## APPENDIX A

The reasons for different behavior of various types of impurities in superconductors, as well as the effect of pairing symmetry, can be illuminated by rewriting the problem in terms of the elementary quasiparticle excitations of the clean superconductor, described by the Hamiltonian, Eq. (6). This Hamiltonian is diagonalized by the Bogoliubov-Valatin transformation, $\Psi_{\mathbf{k}} \to \Gamma_{\mathbf{k}} = S_{\mathbf{k}} \Psi_{\mathbf{k}}$, where $\Gamma_{\mathbf{k}} = (\gamma_{\mathbf{k}\uparrow} \; \gamma^{\dagger}_{-\mathbf{k}\downarrow})^T$ and

$$S_{\mathbf{k}} = \begin{pmatrix} \cos\theta_{\mathbf{k}} & -\sin\theta_{\mathbf{k}} \\ \sin\theta_{\mathbf{k}} & \cos\theta_{\mathbf{k}} \end{pmatrix}. \tag{A1}$$

In this parametrization, $u_{\mathbf{k}} = \cos\theta_{\mathbf{k}}$ and $v_{\mathbf{k}} = \sin\theta_{\mathbf{k}}$. Choosing $\sin 2\theta_{\mathbf{k}} = \Delta_{\mathbf{k}}/E_{\mathbf{k}}$ gives

$$H_0 = \sum_{\mathbf{k}} E_{\mathbf{k}} \Gamma^{\dagger}_{\mathbf{k}} \hat{\tau}_3 \Gamma_{\mathbf{k}}, \tag{A2}$$

where $E_{\mathbf{k}} = \sqrt{\epsilon_{\mathbf{k}}^2 + \Delta_{\mathbf{k}}^2}$. In terms of these excitations, the Hamiltonian for the $\delta$-function impurity is $H_{\text{imp}} = \sum_a H^{(a)}$, with

$$H^{(a)} = \frac{1}{N} \sum_{\mathbf{k}\mathbf{k}'} \Gamma^{\dagger}_{\mathbf{k}} \hat{V}^{(a)}_{\mathbf{k}\mathbf{k}'} \Gamma_{\mathbf{k}'}. \tag{A3}$$

The three types of scattering interactions ($a = w, u, \Delta$) are: (i) magnetic,

$$\hat{V}^{(w)}_{\mathbf{k}\mathbf{k}'} = w[\cos(\theta_{\mathbf{k}} - \theta_{\mathbf{k}'})\hat{\tau}_0 - \sin(\theta_{\mathbf{k}} - \theta_{\mathbf{k}'})i\hat{\tau}_2]; \tag{A4}$$

(ii) potential,

$$\hat{V}^{(u)}_{\mathbf{k}\mathbf{k}'} = U[\cos(\theta_{\mathbf{k}} + \theta_{\mathbf{k}'})\hat{\tau}_3 + \sin(\theta_{\mathbf{k}} + \theta_{\mathbf{k}'})\hat{\tau}_1]; \tag{A5}$$

and (iii) local order-parameter suppression,

$$\hat{V}^{(\Delta)}_{\mathbf{k}\mathbf{k}'} = -\delta\Delta[\sin(\theta_{\mathbf{k}} + \theta_{\mathbf{k}'})\hat{\tau}_3 - \cos(\theta_{\mathbf{k}} + \theta_{\mathbf{k}'})\hat{\tau}_1], \tag{A6}$$

where $\delta\Delta \equiv \Delta_0 - \Delta(\mathbf{r} = 0) \geq 0$. These expressions form the basis for the following discussion of when to expect a bound state in the superconducting energy gap.

### 1. s-wave superconductor

First note that, for $k \sim k_F$, $\theta_{\mathbf{k}} \sim \frac{\pi}{4}\text{sgn}(\Delta_{\mathbf{k}})$. For an s-wave superconductor at energies close to the chemical potential, the magnetic interaction term is diagonal: $\hat{V}^{(w)}_{\mathbf{k}\mathbf{k}'} \sim w\hat{\tau}_0$. Thus, the excitation energy is locally decreased for spin-down excitations and increased for spin-up excitations (recall that the effective $w$ changes sign for opposite spins of quasiparticles). As a consequence, a bound state appears only for the spin-down quantum number [32]. For large enough $w$, the bound-state energy crosses the chemical potential.

The potential scattering term is purely off-diagonal: $\hat{V}^{(u)}_{\mathbf{k}\mathbf{k}'} \sim U\hat{\tau}_1$; it causes level repulsion regardless of the sign of $U$. Therefore, potential scattering does not introduce intragap states.

Local suppression of the order parameter leads to an attractive diagonal potential for both type of quasiparticles: $\hat{V}^{(\Delta)}_{\mathbf{k}\mathbf{k}'} \sim -\delta\Delta\hat{\tau}_3$. This produces two degenerate bound states in the energy gap, as expected from time-reversal symmetry.

### 2. d-wave superconductor

In contrast to the s-wave case, the alternating sign of the d-wave order parameter yields quasiparticle scattering terms that are predominately neither diagonal nor off-diagonal. While a large enough potential will lead to pronounced resonance states, they do not cross the chemical potential when the normal state quasiparticle spectrum has particle-hole symmetry and the gap function average over the Fermi surface vanishes.

Finally, the appearance of degeneracies in the spectrum of virtual bound states can be understood in terms of the symmetry properties of the effective Hamiltonian, $H = H_0 + H_{\text{imp}}$. In general, a d-wave superconductor is invariant under the transformation in which $\Psi_{\mathbf{k}} \to \Psi'_{\mathbf{k}} = \hat{\tau}_3 \Psi_{R\mathbf{k}}$, where $R$ denotes a rotation through an angle $\frac{\pi}{2}$ about the impurity site. When the system is half filled and $U = 0$, the symmetry group of $H$ includes the charge-conjugation transformation. Because such a symmetry group is non-Abelian, it also has higher than one-dimensional irreducible representations. In particular, the impurity-induced virtual bound states belong to a two-dimensional irreducible representation; i.e., they are two-fold degenerate. The degeneracy is lifted by a nonzero $U$, by an asymmetric band structure, or by a gap function which has a nonzero s-wave component. For $w = 0$, the quasiparticle states are at least two-fold degenerate because of time-reversal symmetry.

### APPENDIX B

In this Appendix, a mapping to an effective theory is presented, from which the consequences of a finite density of impurities can be explored. To formulate the problem of many magnetic impurities, we utilize the four-dimensional Gor'kov-Nambu representation, because the magnetic moments are allowed to be randomly oriented. In this representation, the effective Hamiltonian is $H = H_0 + H_{\text{imp}}$, where

$$H_0 = \sum_{\mathbf{k}} \Psi^{\dagger}_{\mathbf{k}}(\epsilon_{\mathbf{k}}\hat{\tau}_3 + \Delta_{\mathbf{k}}\hat{\tau}_2\hat{\sigma}_2)\Psi_{\mathbf{k}}, \tag{B1}$$

and $\Psi_{\mathbf{k}} = (\psi_{\mathbf{k}\uparrow} \; \psi_{-\mathbf{k}\downarrow} \; \psi^{\dagger}_{\mathbf{k}\uparrow} \; \psi^{\dagger}_{-\mathbf{k}\downarrow})^T$ is the Gor'kov-Nambu spinor, $\hat{\tau}_\alpha$ and $\hat{\sigma}_\alpha$ ($\alpha = 1, 2, 3$) are the Pauli matrices for particle-hole and spin degrees of freedom, and $\hat{\tau}_0$ and $\hat{\sigma}_0$ are the unit matrices. We further assume that randomly distributed impurities scatter by $\delta$-function interactions,

$$H_{\text{imp}} = \sum_{\mathbf{r}} \Psi^{\dagger}(\mathbf{r})\hat{V}(\mathbf{r})\Psi(\mathbf{r}) \tag{B2}$$

where

$$\hat{V}(\mathbf{r}) = \sum_{n} \hat{v}_n \delta_{\mathbf{r}\mathbf{r}_n}. \tag{B3}$$

Here, $\hat{v}_n = \mathbf{w}_n \cdot \hat{\mathbf{e}} + U_n \hat{\tau}_3$ is the interaction strength of the $n$th impurity moment located at $\mathbf{r}_n$. Following the usual convention, we have defined $\hat{\mathbf{e}} = \hat{\sigma}_2 \mathbf{e}_2 + \hat{\tau}_3(\hat{\sigma}_1 \mathbf{e}_1 + \hat{\sigma}_3 \mathbf{e}_3)$. In the absence of the impurities, the single-particle Green's function is

$$\hat{G}^{(0)}(\mathbf{r}, \omega) = \frac{1}{N} \sum_{\mathbf{k}} \frac{e^{i\mathbf{k} \cdot \mathbf{r}}}{\omega - \epsilon_{\mathbf{k}} \hat{\tau}_3 - \Delta_{\mathbf{k}} \hat{\tau}_2 \hat{\sigma}_2 + i0^+}. \quad (B4)$$

Generally, the total Green's function in the presence of impurities is given by

$$\hat{G}(x, x') = \hat{G}^{(0)}(x - x') + \int dy dy' \, \hat{G}^{(0)}(x - y) \hat{T}(y, y') \hat{G}^{(0)}(y' - x'), \quad (B5)$$

where $\hat{T}$ is a $T$ matrix. We use the notation in which all the matrices are given in four-dimensional Gor'kov-Nambu space, $x = (\mathbf{r}, t)$, and $\int dx = \sum_{\mathbf{r}} \int dt$. The $T$ matrix is a solution of the Lippmann-Schwinger equation,

$$\hat{T}(y, y') = \hat{V}(y)\delta(y - y') + \int dz \, \hat{V}(y) \hat{G}^{(0)}(y - z) \hat{T}(z, y'). \quad (B6)$$

For impurities with $\delta$-function interactions, it is straightforward to see that a $T$ matrix of the form

$$\hat{T}(\mathbf{r}, \mathbf{r}'; \omega) = \sum_{mn} \hat{T}_{mn}(\omega) \delta_{\mathbf{r}\mathbf{r}_m} \delta_{\mathbf{r}'\mathbf{r}_n}, \quad (B7)$$

solves the Lippmann-Schwinger equation. Indeed, defining

$$\begin{aligned} \langle m | \mathbf{T}(\omega) | n \rangle &= \hat{T}_{mn}(\omega), \\ \langle m | \mathbf{G}^{(0)}(\omega) | n \rangle &= \hat{G}^{(0)}(\mathbf{r}_m - \mathbf{r}_n; \omega), \\ \langle m | \mathbf{v}^{-1} | n \rangle &= \hat{v}_m^{-1} \delta_{mn}; \end{aligned} \quad (B8)$$

the solution is

$$\mathbf{T}(\omega) = [\mathbf{v}^{-1} - \mathbf{G}^{(0)}(\omega)]^{-1}. \quad (B9)$$

This formula allows us to compute various physical properties of the system in the case of a finite number of random impurities for $s$- or $d$-wave superconductors. One such quantity of interest is the local spectral density. Eq. (B9) also implies a useful correspondence between the current problem and that of noninteracting quasiparticles moving on an infinite lattice with random on-site energies and hopping amplitudes (see below).

Consider the local spectral density,

$$A_\sigma(\mathbf{r}, \omega) = -\frac{1}{\pi} \text{Im} \, G_{\sigma\sigma}(\mathbf{r}, \mathbf{r}; \omega), \quad (B10)$$

and define $A_\sigma(\mathbf{r}, \omega) = A_0(\omega) + \delta A_\sigma(\mathbf{r}, \omega)$, where $A_0(\omega) = -\pi^{-1} \text{Im} \, G_{\sigma\sigma}^{(0)}(\mathbf{r} = 0, \omega)$ is the spectral density in a clean superconductor. For example, for an $s$-wave superconductor, $A_0(\omega) = N_F |\omega| / \sqrt{\omega^2 - \Delta_0^2}$, when $|\omega| > \Delta_0$, and zero otherwise. The second term,

$$\delta A_\sigma(\mathbf{r}, \omega) = \quad (B11)$$
$$-\frac{1}{\pi} \text{Im} \sum_{mn} \langle \sigma | \hat{G}^{(0)}(\mathbf{r} - \mathbf{r}_m, \omega) \hat{T}_{mn}(\omega) \hat{G}^{(0)}(\mathbf{r}_n - \mathbf{r}, \omega) | \sigma \rangle,$$

incorporates the effect of impurities. We have adopted the notation, according to which $\langle \sigma | \hat{G} | \sigma \rangle = G_{\sigma\sigma}$, etc. (Note that $G$ is the particle component of $\hat{G}$: $G \equiv \hat{G}_{11}$.) Because the impurities modeled by Eq. (B3) do not change the total number of states, $\int d\omega \, \delta A_\sigma(\mathbf{r}, \omega) = 0$. Eq. (B11) can be written in a compact form, if we consider the spectral density only at a given impurity site $\mathbf{r}_n$:

$$\delta A_\sigma(\mathbf{r}_n, \omega) = -\frac{1}{\pi} \text{Im} \, \langle n\sigma | \mathbf{G}^{(0)}(\omega) \mathbf{T}(\omega) \mathbf{G}^{(0)}(\omega) | n\sigma \rangle. \quad (B12)$$

The importance of Eq. (B12) becomes clear by a following example. Assume that for a given $\omega = \omega_\alpha$ there exists a state $|\varphi_\alpha\rangle$ such that

$$[\mathbf{v}^{-1} - \mathbf{G}^{(0)}(\omega_\alpha)] |\varphi_\alpha\rangle = 0. \quad (B13)$$

Then, averaging over the random orientation of magnetic moments, the spin-unpolarized spectral density becomes

$$\delta A(\mathbf{r}_n, \omega) = C_n \sum_{\alpha\sigma} \langle n\sigma | \varphi_\alpha \rangle \langle \varphi_\alpha | n\sigma \rangle \mathcal{L}_\alpha(\omega), \quad (B14)$$

where $C_n = (w_n^2 + U_n^2)/(w_n^2 - U_n^2)^2$ and the Lorentzian function $\mathcal{L}_\alpha(\omega)$ is given by $\mathcal{L}_\alpha(\omega) = (2\pi)^{-1} f_\alpha \omega_\alpha'' / [(\omega - \omega_\alpha')^2 + \omega_\alpha''^2]$, $\omega_\alpha = \omega_\alpha' + i\omega_\alpha''$, and $f_\alpha$ is the residue of the $\alpha$th pole. This form is useful when $f_\alpha$ depends only weakly on $\omega$. The state $|\varphi_\alpha\rangle$ specifies the properties of quasiparticles at a given energy $\omega_\alpha$. In particular, the matrix element $|\langle n\sigma | \varphi_\alpha \rangle|^2$ incorporates the spatial information of the impurity-induced states; for example, whether the quasiparticle state at the energy $\omega = \omega_\alpha$ is localized or extended.

The above result suggests a fruitful formulation for exploring impurity-induced quasiparticle states in the superconducting energy gap. The central idea is to define the Hamiltonian,

$$\mathcal{H} = \sum_m c_m^\dagger \hat{v}_m^{-1} c_m - \sum_{mn} c_m^\dagger \hat{t}_{mn} c_n, \quad (B15)$$

where $c_m$ is a four-component spinor and $\hat{t}_{mn} = \hat{G}^{(0)}(\mathbf{r}_m - \mathbf{r}_n; \omega)$. Noting the similarity with Eq. (B13), we can immediately conclude that the zero-energy eigenstate of $\mathcal{H}$ also determines the properties of a quasiparticle state in the impurity band at a given energy $\omega$. For example, in a three-dimensional $s$-wave superconductor where magnetic impurities yield localized states in the energy gap, the formation of the impurity band is mapped to the Hamiltonian $\mathcal{H}$ with short-range hopping amplitudes. Indeed, $\hat{t}_{mn} \propto r^{-1} e^{-r_{mn}/\lambda}$, where $r_{mn} = |\mathbf{r}_m - \mathbf{r}_n|$. In $d$-wave superconductors, Eq. (B15) has been applied to study localization properties of quasiparticle states that are induced by unitary scatterers [31].